\documentclass[useAMS,usenatbib]{mn2e}

\usepackage{txfonts,natbib,graphicx,color}

\bibpunct{(}{)}{;}{a}{}{,} 
\newcommand{\myemail}{papitto@ice.csic.es}

\def\src{XSS J12270--4859}

\def\xmm{{\it XMM-Newton}}
\def\1023{PSR J1023+0038}

\title[X-ray pulsation from {\src}]
  {X-ray coherent pulsations during a sub-luminous accretion disk state of the transitional millisecond pulsar {\src}}
\author[A. Papitto et al.]
  {A.~Papitto$^1$\thanks{\myemail}, 
  D.~de Martino$^2$, T.~M.~Belloni$^3$, M.~Burgay$^4$, A.~Pellizzoni$^4$, A.~Possenti$^4$, \newauthor D.~F.~Torres$^{1,5}$\\
  $^1$Institut de Ci\`encies de l'Espai (IEEC-CSIC), Campus UAB,
  Fac. de Ci\`encies, Torre C5, parell, 2a planta, 08193 Barcelona, Spain\\
  $^2$INAF - Osservatorio Astronomico di Capodimonte, salita Moiariello 16, 80131, Napoli, Italy\\
  $^3$INAF - Osservatorio Astronomico di Brera, via E Bianchi 46, 23807, Merate (LC), Italy\\
  $^4$INAF - Osservatorio Astronomico di Cagliari, Via della Scienza, I-09047 Selargius (CA), Italy\\
  $^5$Instituci\'o Catalana de Recerca i Estudis Avan\c{c}ats (ICREA), E-08010 Barcelona, Spain}
\date{}

\pagerange{\pageref{firstpage}--\pageref{lastpage}} \pubyear{2014}

\def\LaTeX{L\kern-.36em\raise.3ex\hbox{a}\kern-.15em
    T\kern-.1667em\lower.7ex\hbox{E}\kern-.125emX}

\begin{document}

\label{firstpage}

\maketitle

\begin{abstract}
We present the first detection of X-ray coherent pulsations from the
transitional millisecond pulsar {\src}, while it was in a sub-luminous
accretion disk state characterized by a 0.5--10 keV luminosity of
$5\times10^{33}$ erg s$^{-1}$ (assuming a distance of 1.4
kpc). Pulsations were observed by {\xmm} at an rms amplitude of
$(7.7\pm0.5)\%$ with a second harmonic stronger than the the
fundamental frequency, and were detected when the source is neither
flaring nor dipping. The most likely interpretation of this detection
is that matter from the accretion disk was channelled by the neutron
star magnetosphere and accreted onto its polar caps. According to
standard disk accretion theory, for pulsations to be observed the mass
in-flow rate in the disk was likely larger than the amount of plasma
actually reaching the neutron star surface; an outflow launched by the
fast rotating magnetosphere then probably took place, in agreement
with the observed broad-band spectral energy distribution.  We also
report about the non-detection of X-ray pulsations during a recent
observation performed while the source behaved as a
rotationally-powered radio pulsar.
\end{abstract}


\section{Introduction}

The extremely short spin periods of millisecond pulsars are the
outcome of a Gyr-long phase of accretion of mass transferred by a low
mass ($< M_{\odot}$) companion star through an accretion disk
\citep{alpar1982,radhakrishnan1982}. During the mass accretion phase
these systems are bright X-ray sources. When mass transfer eventually
declines, a pulsar powered by rotation of its magnetic field turns on,
emitting from the radio to the gamma-ray band. The $\sim$300
millisecond radio pulsars (MSP) in our Galaxy are then believed to be
the recycled descendants of accreting neutron stars (NS) in low mass
X-ray binaries (NS-LMXB). Indeed, accretion-powered pulsations at a
period of few ms were detected from 15 NS-LMXBs, so far
\citep{wijnands1998}, due to the channelling of the mass in-flow to
the magnetic poles of the NS by the magnetosphere. These sources are
dubbed accreting millisecond pulsars (AMSP, see \citealt{patruno2012c}
for a review).

Recently, the tight link between  MSPs and NS-LMXBs has been
highlighted by the discovery of three transitional ms pulsars, sources
that switched between accretion and rotation-powered emission on time
scales ranging from a few weeks to a few years. These include, (i) IGR
J18245--2452, a binary of the globular cluster M28 that turned on as a
bright ($L_X\approx 10^{36}$ erg s$^{-1}$) AMSP in 2013, and was
observed as a rotationally-powered MSP a few years before, and a few
weeks after the accretion event \citep{papitto2013nat}; (ii) PSR
J1023+0038, an MSP that had a sub-luminous ($\la10^{34}$ erg s$^{-1}$)
accretion disk in 2000/2001 \citep{archibald2009}, and that have
entered back again in such a state in 2013
\citep{patruno2014,stappers2014}; (iii) {\src}, an LMXB that remained
for a decade in a sub-luminous disk accretion phase
\citep{saitou2009,demartino2010,demartino2013}, characterized by
correlated X-ray and UV flux variability \citep{demartino2013} and by
bright radio and GeV emission \citep{demartino2010,hill2011}; during
December 2012 it then transited to an MSP state characterized by the
detection of 1.69 ms radio pulsations \citep{roy2014paper}, a fainter X-ray
and gamma-ray emission
\citep{tam2013,bassa2014,bogdanov2014,xing2014}, and the absence of an
accretion disk \citep{demartino2014}.

While only IGR J18245--2452 has been observed in a bright X-ray
outburst so far, all the three transitional ms pulsars known showed a
sub-luminous disk state \citep[see, e.g.][]{linares2014}. Such
a state is characterized by (i) an accretion disk around the NS; (ii)
a highly variable X-ray emission at a level of
$\mbox{few}\times10^{33}$ erg s$^{-1}$, intermediate between the
luminosity shown by X-ray transients in outburst ($\ga 10^{36}$ erg
s$^{-1}$) and in quiescence ($\la 10^{32}$ erg s$^{-1}$); (iii) a
gamma-ray ($>0.1$ GeV) luminosity of $\mbox{few}\,\times\,10^{33}$ erg
s$^{-1}$, from two- to ten-times brighter with respect to the emission
during the radio pulsar phase\footnote{Note that gamma-ray emission
  from IGR J18245--2452 cannot be resolved as it belongs to a globular
  cluster}; (iv) a radio continuum emission characterized by a flat
spectrum, typical of outflows launched by compact objects in
LMXBs. Such a complex phenomenlogy was interpreted in a number of
studies in terms of an enshrouded radio MSP turned on in spite of the
presence of the disk, producing high-energy radiation at the shock
between the pulsar wind and the in-flowing matter
\citep{stappers2014,takata2014,cotizelati2014,li2014}. On the other
hand, \citet{papitto2014} and Papitto \& Torres (2014, submitted)
proposed that matter penetrated inside the light cylinder turning off
the rotationally-powered pulsar, while the system ejects matter from the
inner disk boundary as a propeller.

Here we present an analysis of observations of {\src} performed by
{\xmm} in January 2011 and June 2014, when the source was in the
sub-luminous disk state and in the rotationally-powered state,
respectively. This analysis was aimed at searching for a coherent
signal by making use of the recently obtained radio pulsar ephemeris
(Burgay et al. in prep.) derived from observations that have been
performed during the rotationally-powered state in which the source is
found since December 2012.

\section{Observations}

 {\src} was observed by {\xmm} four times between 2009 and 2014.  To
 search for a signal at the 1.69 ms spin period, we considered only
 the observations performed with the EPIC-pn camera operated in a fast
 timing mode with a time resolution of 29.5 $\mu$s, i.e. those
 performed on 2011 Jan 01 (Obs. Id 0656780901) and 2014 Jun 27
 (Obs. Id 0729560801). In the first one the source was in a
 sub-luminous disk state, while in the latter the source behaved as a
 rotationally-powered radio MSP. During these observations the EPIC
 cameras were equipped with a thin optical blocking filter. Periods of
 high flaring background were identified during the 2014 observation
 and removed from the analysis, reducing the effective exposure to
 39.4 ks, while the whole 30 ks exposure of the 2011 observation was
 retained. We analyzed data using the {\xmm} Science Analysis
 Software\footnote{http://xmm.esac.esa.int/sas/}, v. 14.0.

 Source photons observed during the 2011 observation were extracted
 from an 86.1"-wide strip around the source position (equivalent to 21
 pixels and enclosing 98\% of the energy), while the background was
 estimated from a strip of 12" of width, far from the source. As
 during the 2014 observation the source emission was dominated by
 background, we considered a smaller 45.1"-wide strip (enclosing 93\%
 of the energy). In order to estimate the source flux during this
 observation we considered data taken by the MOS cameras, which were
 operated in full window mode, thus retaining their imaging
 capabilities; a circular region of 50" of radius around the source
 position was considered to include 90\% of the photons emitted by the
 source, while the background was extracted from a 115" circular
 region without any source.

X-ray photons were preliminary reported to the Solar System barycentre
using the position of the optical counterpart of {\src} determined by
\citet{masetti2006}, RA=12$^{h}$ 27$^{m}$ 58.748$^{s}$, DEC=$-48^{\circ}$
53' 42.88".

\section{Analysis and results}

\subsection{2011 observation}

\begin{figure}
{\includegraphics[angle=0,width=\columnwidth]{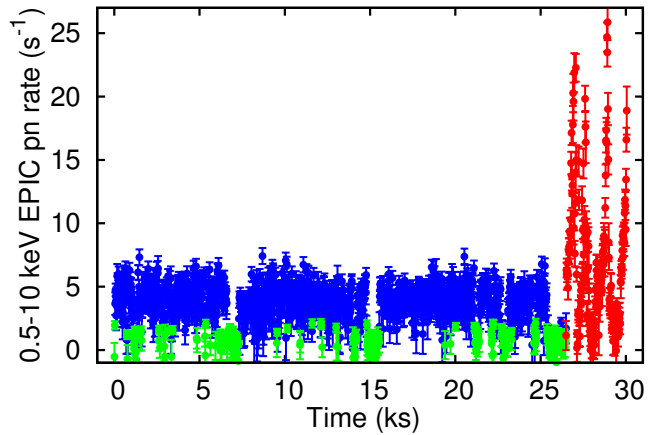}}
\caption{Background subtracted, 0.5-10 keV EPIC pn light curve
  observed on 2011 Jan 01. Quiescent, dip, and flaring state (see text
  for the definition) are plotted in blue, green and red,
  respectively.}
 \label{fig:lcurve}
\end{figure}

The light curve observed by the EPIC pn during the 2011 observation is
plotted in Fig.~\ref{fig:lcurve}. The source showed flaring activity
during the last 4 ks of the observation (red points in
Fig.~\ref{fig:lcurve}), while during the remaining of the observation
it was mostly found at a net countrate of $\sim 4.5$ s$^{-1}$ (steady
quiescent state according to the terminology used in
\citealt{demartino2013}, see blue points in
Fig.~\ref{fig:lcurve}). This corresponds to an unabsorbed 0.5-10 keV
flux of $2.1\times10^{-11}$ erg cm$^{-2}$ s$^{-1}$. Dips were also
observed (green points in Fig.~\ref{fig:lcurve}), and a threshold of $2$
s$^{-1}$ on the net count rate was used to distinguish between
quiescent and dip emission. For details on the light curve and
spectrum the reader is referred to \citet{demartino2013}.

\begin{figure}
 \resizebox{\hsize}{!} {\includegraphics{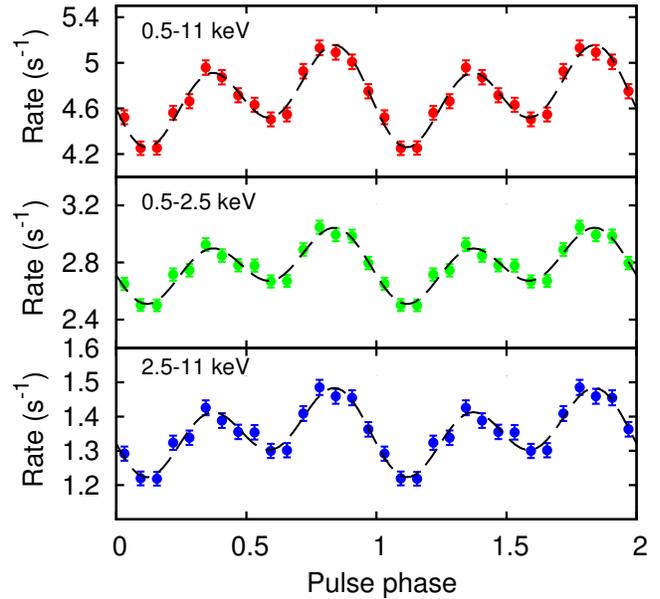}}
\caption{Background subtracted pulse profile observed during the
  steady quiescent state of the 2011 {\xmm} observation in the 0.5-11 keV
  (top panel), 0.5-2.5 keV (middle panel), 2.5-11 keV (bottom panel)
  energy bands. Two cycles are shown for clarity.  }
 \label{fig:profile}
\end{figure}

\begin{figure}
 \resizebox{\hsize}{!} {\includegraphics{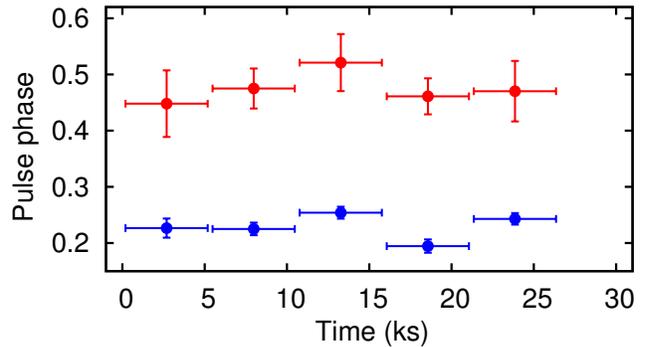}}
\caption{Phase of the first (red points, top) and second harmonic
  (blue points, bottom) observed during the steady quiescent state.  }
 \label{fig:phase}
\end{figure}

In order to search for pulsations we used the pulsar ephemeris derived
from a timing analysis of the radio pulsed signal detected during a
series of oservations performed with the Parkes 64-m antenna in 2014
(Burgay et al., in prep.; see central column of Table~\ref{tab}), and
obtained assuming a circular orbit. Even if the spin-down rate of
{\src} is not known, the typical rates observed from similar MSP
($\mbox{few}\times10^{-15}$ Hz s$^{-1}$) ensured that only a single
frequency had to be searched for pulsations in 2011 data.
\citet{caliandro2012} estimated the amount of power lost $\epsilon$
when folding data with orbital parameters that are different than the
actual ones:
\begin{eqnarray}
\delta{(a\sin{i}/c)}&=&\frac{1}{2\pi\nu}\frac{1}{\epsilon^2}\\ 
 \delta{T^*}&=&\frac{0.1025  P_{orb}}{\pi\nu (a\sin{i}/c)}\frac{1}{\epsilon^2}\\ 
\delta{P_{orb}}&=&\frac{{P_{orb}}^2}{2\pi\nu
  (a\sin{i}/c) \Delta T} \sqrt{\left(\frac{1-\epsilon^2}{10}\right)}.
\end{eqnarray}
Here, $\nu$ is the NS spin frequency, $a\sin{i}/c$ is the projected
semi-major axis of the NS orbit, $P_{orb}$ is the orbital period and
$T^*$ is the epoch of the NS passage at the ascending node of the
orbit. Considering the accuracy of the parameters of the radio timing
solution, and setting $\epsilon=0.8$, we
concluded that only a search over plausible values of the epoch of
passage at the ascending node of the orbit $T^*$ (in steps of 3.2 s to
cover an interval of 945 s) had to be performed. By folding X-ray data
around the radio ephemeris and performing an epoch folding search, we
found a coherent signal that had a statistical significance of
15-$\sigma$ after taking into account the number of trials made. The
values we measured for the spin frequency and the epoch of passage at
the ascending node are given in the rightmost column of
Table~\ref{tab}; uncertainties were evaluated following
\cite{leahy1987}.  During the steady quiescent state the signal had at a root
mean-squared amplitude of $A_{rms}=(7.7\pm0.5)\%$ (corrected for the
background), and was modelled by two harmonic components, with the
second harmonic stronger by a factor $\simeq1.8$ than the fundamental
frequency (see top panel Fig.~\ref{fig:profile}). The shape of the
pulse profile was similar in a soft (0.5-2.5 keV) and a hard (2.5-11
keV) energy band, with rms amplitude of $(6.1\pm0.9)\%$ and
$(7.5\pm0.7)\%$, respectively (see middle and bottom panel of
Fig.~\ref{fig:profile}).  The phase of the second harmonic varied over
intervals of 5 ks while the phase of the first one was stable within
the errors, indicating slight changes of the pulse profile during the
steady quiescent state (see Fig.~\ref{fig:phase}); no residual variability at
the orbital period was found.

The variance of the pulse profile obtained by folding the X-ray
photons observed during the dipping and the flaring activity has a
probability of being produced by photon counting noise of 5.8 and
2.7$\%$, respectively. We then set upper limits on the background
corrected rms amplitude observed during the dipping and the flaring
state of 5.9 and 2.0$\%$ (3-$\sigma$ confidence level), respectively.
If pulsations were present during these states, their amplitude was
then lower than during the steady quiescent state.

\begin{table}
\caption{Spin and orbital parameters of {\src}}
\label{tab}
\centering
\renewcommand{\footnoterule}{} 
\begin{tabular}{@{}l c c}

\hline
Parameter & {\it PKS} (2014) & {\it XMM} (2011) \\
\hline
$\nu$ (Hz) & 592.98777209(84) & 592.9877712(22) \\
$P_{orb}$ (s) & 24874.27(38) & ... \\
$a\sin{i}/c$ (lt-s) & 0.668504(17) & ... \\
$T^*$ (MJD) & 56718.1766(18) & 55562.3121504(23)\\
$T_{ref}$ (MJD) &  56718.39814 & 55562.296372\\
\hline
\hline
\end{tabular}
\end{table}

\subsection{2014 observation}

During the 2014 observation {\src} was in a rotationally-powered
state. It was much fainter in X-rays than in 2011, as it could not be
detected at its position in the one-dimensional image of the EPIC-pn
chip due to the contribution of close-by sources. To evaluate the
source flux we extracted a spectrum from data taken by the two MOS
cameras. The spectrum in the 0.3-10 keV energy range was modelled with
a power law with an index of $\Gamma=1.07(7)$, giving a 0.5-10 keV
flux of $7.0(5)\times10^{-13}$ erg cm$^{-2}$ s$^{-1}$. We evaluated
with
webpimms\footnote{https://heasarc.gsfc.nasa.gov/cgi-bin/Tools/w3pimms/w3pimms.pl}
the countrate expected for the EPIC pn as 0.15 s$^{-1}$. Such a
count-rate gives an expected signal-to-noise ratio of unity in the
45.1"-wide stripe of the EPIC pn chip that we used to extract
photons. Folding the observed X-ray photons around the radio pulsar
ephemeris, and performing a search on possible values of the epoch of
passage at the ascending node analogous to that carried out on 2011
data, did not result in significant detection. We set a upper limit on
the background subtracted rms amplitude of $7.1\%$ (3-$\sigma$
confidence level).  The signal was not detected even in the 0.5-2.5
keV range in which X-ray pulsations were detected by
\citet{archibald2009} from the twin MSP PSR J1023+0038. We set an
upper limit at a 3-$\sigma$ confidence level on the rms amplitude of
$9.8\%$. Similar results were obtained when searching for a signal at
the second harmonic of the signal.  During the observation of June
2014, {\src} showed an orbital modulation with an amplitude of
$\approx 70\%$, similar to that already observed by
\citet{bogdanov2014} in an {\xmm} observation performed in December
2013. While a detailed analysis of the orbital characteristics will be
presented elsewhere, a search for pulsations restricted to orbital
phases close to superior conjunction (i.e., when the pulsar
contribution to the X-ray emission is expected to be larger) gave an
upper limit of 14\% on the rms amplitude.

\section{Discussion}

We reported the first detection of X-ray pulsation at the 1.69 ms spin
period from {\src} while it was in a sub-luminous disk
state. Considering the enigmatic nature of such a state, both a
rotationally and an accretion-powered origin are considered next.

\subsection{Rotationally-powered pulsations}

The spin down power of {\src} measured by \citet{roy2014paper}
  during the rotationally-powered MSP state is $0.9\times10^{35}$ erg
  s$^{-1}$. A fraction equal to $10^{-3}$ of this power is converted
  into observed 0.5-10 keV X-rays, similar to other rotationally-powered MSPs
  \citep{possenti2002}.  If {\src} were rotationally-powered also
  during the sub-luminous accretion disk state, an X-ray luminosity of
  $\approx 10^{32}$ erg s$^{-1}$ would be then expected. The
 0.5-10 keV luminosity observed in 2011 is instead much
larger, $L_X\simeq5\times10^{33}$ erg s$^{-1}$ (assuming a distance of
1.4 kpc, as estimated by \citealt{roy2014paper} from the dispersion
measure of radio pulses). Even if only the pulsed luminosity,
$\sqrt{2}A_{rms}L_x \simeq5\times10^{32}$ erg s$^{-1}$, had a
magnetospheric origin, {\src} would still have been five times
overluminous in X-rays than during the rotationally-powered MSP state.
Furthermore, we set un upper limit on the 0.5-10 keV pulsed luminosity
during the 2014 rotationally-powered pulsar state of
$\simeq1.6\times10^{31}$ erg cm$^{-2}$ s$^{-1}$, which is $\sim30$
times smaller than the value observed during the 2011 sub-luminous
disk state. If pulsations observed during the sub-luminous disk state
were of magnetospheric origin, one would not expect such a strong
variation of the X-ray pulsed flux when the source switches to a
purely rotationally-powered state. Together with the previous
energetic considerations, this disfavours an interpretation of the
pulses being produced by a rotationally-powered MSP.

\subsection{Accretion-powered pulsations}

The X-ray pulsations observed from {\src}  closely resemble those
observed from AMSPs, which have an amplitude of few per cent, and are
modelled with two harmonics \citep[see][and references
  therein]{patruno2012}. The  second harmonic of the pulse
observed from {\src} is stronger than  the fundamental
frequency; a similar  shape was observed from the pulses of the
eclipsing AMSP, SWIFT J1749.4--2807
\citep{altamirano2011,ferrigno2011}. According to analytical
calculations made by \citet{poutanen2006b}, the ratio of  the
amplitude of the second harmonic to the fundamental observed
from a fast pulsar whose antipodal spots are always visibile, is
$c_2/c_1=1/2 (\tan{i}\tan{\theta})$, where $i$ is the binary inclination
and $\theta$ is the spot co-latitude. \citet{demartino2014}
constrained the inclination of {\src}  between 45 and 65$^{\circ}$
from the modelling of the optical light curve observed during the
rotationally-powered state. The large ratio observed in the case of
{\src}, $c_2/c_1\simeq1.8$ then  indicates a large spot
co-latitude, $\theta\ga 60^{\circ}$. 

X-ray pulsations were observed from {\src} at a 0.5-10 keV luminosity
of $\simeq5\times10^{33}$ erg s$^{-1}$, more than an order of
magnitude lower than the level at which pulses have been observed from
AMSPs so far ($\approx10^{35}$ erg s$^{-1}$,
\citealt{patruno2009}). Accretion onto the NS surface occurrs
unhindered as long as the disk is truncated within the co-rotation
radius, $R_{co}=23.7\, m_{1.4}^{1/3}$ km for {\src}, where $m_{1.4}$
is the NS mass in units of 1.4 M$_{\odot}$. The radius at which the
magnetosphere truncates the disk is expressed as a fraction
$\xi=0.5-1$ of the Alfven radius \citep[see, e.g.][]{ghosh2007},
$R_{in}\simeq116\,\xi\,\dot{m}_{14}^{-2/7}\,m_{1.4}^{-1/7}\,\mu_{26}^{4/7}$
km, where $\dot{m}_{14}$ is the disk mass in-flow rate in units of
$10^{14}$ g s$^{-1}$, and $\mu_{26}$ is the NS dipole magnetic moment
in units of $10^{26}$ G cm$^{-3}$. We estimate $\mu_{26} = 0.8$
  for {\src} by using the relation given by \citet{spitkovsky2006} to
  derive the NS dipole magnetic moment from the spin down power, and
  considering the limiting case of a purely orthogonal rotator (i.e.,
  $\theta=90^{\circ}$; larger values of $\mu$ are obtained for a
  smaller magnetic inclination angle). In the 10-100 keV band {\src}
emitted a luminosity comparable to that observed in the 0.5-10 keV
band \citep{demartino2010}, giving a bolometric X-ray luminosity of
$L_{X}\simeq10^{34}$ erg s$^{-1}$. Assuming that the whole accretion
power is converted into observable X-ray emission, we then estimate
that while X-ray pulsations were observed,
$\dot{m}_{14}\simeq0.5$. Considering such a mass accretion rate,
  the disk should have been truncated at $R_{in}\ga 60$ km (evaluated
  for $\xi=0.5$ and $m_{1.4}=1$), approximately three times larger
  than the co-rotation radius. Accretion onto the NS surface should
  have been then completely inhibited by the centrifugal barrier set
  by the quickly rotating magnetosphere of the NS, constrasting with
  the observations of accretion-driven X-ray pulsations.

\subsection{An evidence of mass outflow?}

A disk mass accretion rate larger than the
  value deduced from the observed X-ray luminosity is an intriguing
  possibility to reconcile the observation of X-ray pulsation with
  standard disk accretion theory. A larger disk inward pressure would
  in-fact allow the inner disk radius to lie close to the co-rotation
  surface. To satisfy $R_{in}=R_{co}$, a mass accretion rate
  $\dot{m}_{14}$ ranging between 15 and 160 (for $\xi$ in the range
  0.5--1 and $m_{1.4}=1$) is needed.  A similar accretion rate is
  larger by a factor $\ga 30$ than that implied by the X-ray
  luminosity. More than 95\% of the in-flowing disk mass should be
then ejected by the system close to the magnetospheric boundary. 

The possibility that an outflow is launched by {\src} in the
sub-luminous disk state is compatible with the flat (or slightly
inverted) bright radio emission \citep{hill2011}, which is typical of
outflowing LMXBs. Outflows can be launched by a fast rotating
millisecond pulsar due to the propeller effect \citep{illarionov1975}.
\citet{papitto2014} applied a similar scenario to explain the X-ray
and gamma-ray emission observed from {\src} in the disk sub-luminous
state in terms of synchrotron and self-synchrotron Compton emission
emitted at the disk-magnetosphere boundary.  We note that {\src} (as
later also PSR J1023+0038, see below), is the first confirmed
accreting NS to show a bright gamma-ray output.  

If the disk mass in-flow rate is more than 30 times larger than the
rate indicated by the X-ray luminosity, it also follows that the X-ray
radiative efficiency of the accretion disk (which emits half of the
accretion power liberated down to that distance) should be of less
than $20$ per cent, in order to match the observed X-ray
luminosity. Such a low disk X-ray radiative efficiency is of the order
of that recently estimated by \citet{dangelo2014} for Cen
X-4. Accretion disks are expected to become radiatively inefficient as
soon as the mass in-flow rate drops below $\dot{m}_{14}\approx 500$
\citep[e.g.][]{done2007}, a value compatible with the that observed
from {\src}.

According to the so-called radio-ejection scenario
\citep{burderi2001}, the pressure of the rotating NS magneto-dipole is
also a possible driver of ejection of mass from a system hosting a
MSP. Since the observation of X-ray pulsations is a strong indication
that mass accretes onto the NS surface, applying the radio-ejection
scenario to the case of {\src} requires the assumption that the
pressure of the magneto-dipole radiation is able to eject matter even
if a significant fraction of the matter in-flow manages to enter the
light cylinder. This could be the case if the pressure exerted by the
magneto-dipole radiation is not isotropic (e.g. flowing preferentially
along the magnetic equatorial plane), and the magnetic axis of the
dipole is significantly offset with respect to the spin axis (and then
lies close to the disk orbital plane), as indicated by the very strong
second harmonic seen in the pulse profile.

\subsection{A comparison with PSR J1023+0038}

During the preparation of this manuscript, \citet{archibald2014}
reported the detection of X-ray pulsations from PSR J1023+0038 in a
similar sub-luminous disk accretion state as the one in which we
detected pulsations from {\src}. Pulsations were detected only during
quiescent emission (which they dubbed {\it high state}), while not
during dips ({\it low state} according to their terminology), nor
during flares. They also interpreted X-ray pulsations in terms of
channelled accretion onto the NS surface. 

The detection of accretion-powered pulsations from both these systems
rules out the possibility that a radio MSP was active during the
sub-luminous disk state. If it were the case the accretion disk should
have been truncated beyond the light cylinder (80km for these two
sources), which is larger than the corotation radius, thus preventing
accretion onto the NS surface. The X-ray luminosity of PSR J1023+0038
when its showed pulsations was similar to that shown by {\src} and
similar considerations can be made on the value of the disk mass
accretion rate needed to keep the disk inner boundary within the
co-rotation surface, and on the occurrence of outflows. Quite
interestingly, also the pulse profile shown by PSR J1023+0038 in the
sub-luminous disk state is remarkably similar to that shown by
{\src}. The inclination of PSR J1023+0038 is relatively low, $\la
55^{\circ}$ \citep{thorstensen2005,wang2009}, and also in that case a
large spot co-latitude is requested to yield a comparable power in the
first and the second harmonic of the signal.  It remains to be
understood whether a magnetic field configuration with spots close to
the equator, which seems to be relevant to both {\src} and PSR
J1023+0038, might influence the uncommon properties that they
showed in the sub-luminous disk state.

\section*{acknowledgements}

AP and DFT acknowledge support from the the grants AYA2012-39303 and
SGR 2014-1073. AP is supported by a Juan de la Cierva fellowship. DdM
acknowledges financial support from ASI/INAF I/037/12/0. TMB
acknowledges support from INAF PRIN 2012-6. Based on observations
obtained with XMM-Newton, an ESA science mission with instruments and
contributions directly funded by ESA Member States and NASA. We
gratefully thank Dr. Norbert Schartel and the ESAC staff for their
help in obtaining the {\xmm} data.

\label{lastpage}

\bibliography{biblio} \bibliographystyle{mn2e}

\end{document}